\documentclass[aps,prl,twocolumn,superscriptaddress,showpacs]{revtex4-1}
\usepackage{amsmath}
\usepackage{amssymb} 
\usepackage{color}
\usepackage{graphicx}
\usepackage{epstopdf}
\usepackage{hyperref}

\usepackage{mathptmx}

\newcommand{\vect}[1]{\vec{#1}}
\newcommand{\mean}[1]{\left \langle #1 \right \rangle}

\newcommand{\tauenc}{\mean{\tau_{\mathrm{enc}}}}
\newcommand{\tauret}{\mean{\tau_{\mathrm{ret}}}}

\newcommand{\tauin}{\tauenc_{\mathrm{in}}}
\newcommand{\tauout}{\tauenc_{\mathrm{out}}}

\newcommand{\meantau}{\mean{\tau}}

\newcommand{\Pin}{\mathbb{P}_{\mathrm{in}}}
\newcommand{\Pout}{\mathbb{P}_{\mathrm{out}}}

\begin{document}
\title{Encounter times in overlapping domains:  application to epidemic spread in a population of territorial animals}

\author{Luca Giuggioli}
\email{luca.giuggioli@bristol.ac.uk}
\affiliation{Bristol Centre for Complexity Sciences, Department of Engineering Mathematics and School of Biological Sciences, University of Bristol, BS8 1TR, Bristol, UK}

\author{Sebastian P\'{e}rez-Becker}
\email{sebastian.perez.becker@gmail.com}

\author{David P.~Sanders}
\email{dpsanders@ciencias.unam.mx}

\affiliation{Departamento de F\'{i}sica, Facultad de Ciencias, and Centro de Ciencias de la Complejidad (C3), Universidad Nacional Aut\'{o}noma de M\'{e}xico, Ciudad Universitaria,  M\'{e}xico D.F. 04510, Mexico}

\begin{abstract}
We develop an analytical method
 to
calculate encounter times of two random walkers in one dimension when each individual
is segregated in its own spatial 
domain and shares with its neighbor only a fraction of the available space,
finding very good agreement 
with numerically-exact calculations. 
We model a population of susceptible and infected territorial individuals with this spatial 
arrangement, and
which may transmit an epidemic when they meet. We apply the results on encounter times to determine analytically the macroscopic propagation speed of the epidemic as a function of the microscopic characteristics: the confining geometry, the animal diffusion constant, and the infection transmission probability.
\end{abstract}



\maketitle

The spatial propagation of an epidemic in a population is highly dependent on the transmission dynamics, 
in particular, the frequency of encounters between individuals, or contact rate, and the probability of transmission upon  
encounter
\cite{diekmannheesterbeekbook2000}. 
These, in turn, are affected by 
individual mobility \cite{brockmann_prx},  by 
spatial structure of the environment \cite{brockmann_pnas}, and by what is transmitted, e.g., an infection, a rumor, or information, which influences the contact network via which the epidemic spreads \cite{barrat_vespignani}.

Recent studies on the importance of individual-level processes on disease propagation have reemphasized the need to develop models 
that go beyond the assumptions of well-mixed, homogeneous populations 
\cite{fergusonetal2003,colizzavespignani2007,apariciopascual2007,balcanvespignani2011},  
and to link
    `macroscopic' features of the spatial spread of an infection to  `microscopic' characteristics of the agents carrying the infection.
    In this Letter, we make a considerable advance in that direction by 
 focusing on the role that the spatial organization of a population plays for the spread of an epidemic through that population.
 We 
relate the   propagation speed of the epidemic to   pairwise interaction events, consisting of direct encounters between one susceptible and one infected individual.

The rate of encounters between individuals has previously been analyzed 
when
different individuals occupy the same spatial region  \cite{sanderslarralde2008, sanders2009,tejedoretal2011}, based upon
 recent theoretical developments on first-passage times \cite{rednerbook2001}  in confined geometries 
\cite{condaminetal2007,benichouetal2010,benichouetal2011b}.
Here,
we develop a framework to study encounter times when the regions occupied by  individuals are distinct.

\paragraph{Model:-}
Individuals are modeled as random walkers  confined to regions
of equal size; a fraction of each region  overlaps those of the neighboring individuals.
This spatial arrangement is  common in territorial animals: individuals are confined within a home range \cite{giuggiolietal2006}, having exclusive access to certain core areas, the territories, \cite{giuggiolietal2011}, but also sharing other regions, the home-range overlap, with their neighbors. 

\begin{figure}[tb]
\noindent\includegraphics[width=0.9\linewidth]{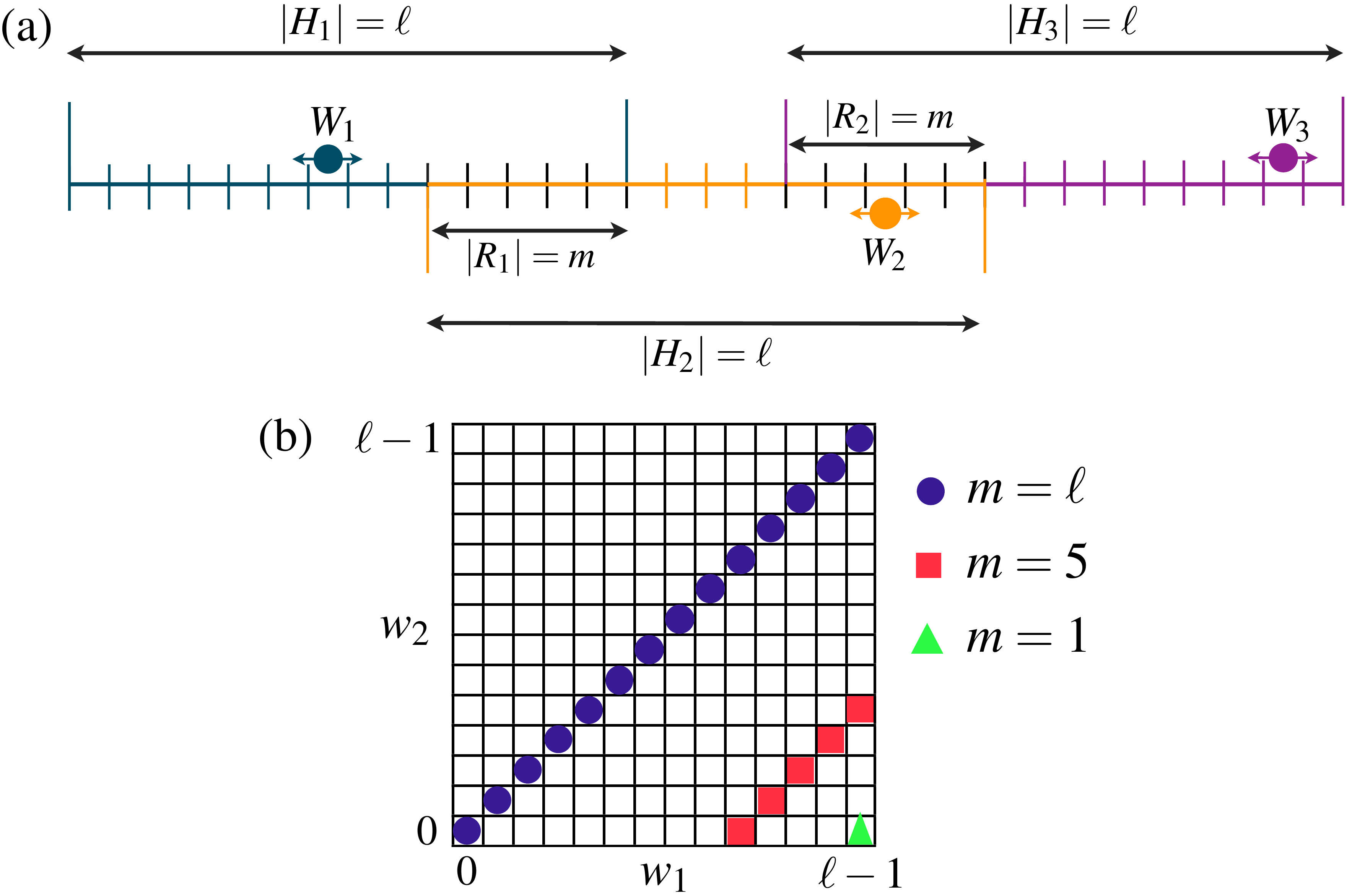}
\caption{(Color online) (a) Three discrete random walkers, with positions 
$W_i$ on a one-dimensional lattice, in their respective habitats $H_i$, 
having overlap regions $R_i$ with their neighbors.
 (b) Mapping of  $W_1$ and $W_2$ to shifted coordinates $(w_{1},w_{2})$ of a single 2D  walker. Static targets corresponding to the locations where the two walkers meet are shown for $m=5$ (squares), $m=1$ (triangles) and $m=\ell$ (circles).}
\label{fig:model}
\end{figure}

When a population is arranged spatially, as depicted in Fig.~\ref{fig:model}, on a one-dimensional (1D) lattice, infection spread proceeds by contact between one infected individual and its neighbor;
encounters may  take place only within the corresponding overlap region. To represent an SI (susceptible--infected) model, we suppose that, upon encounter, transmission of the infection 
takes place with probability  $p$.
 
The discrete position of the $i$th individual is denoted by $W_i$; it is restricted to
its home range region $H_i$ by reflecting boundaries, and has overlap regions  $R_{i-1}$ and $R_{i}$ with, respectively, its left and right neighbor.
The sizes of both types of region are constant:
the $H_{i}$ each occupy $\ell$ sites, and the $R_i$  occupy $m$ sites. 
A similar geometry has been used independently in a model for  heat conduction 
 in 
1D systems  
\cite{ryals_young_2012, huveneers_2011}.

\paragraph{Encounter times:-}
We first suppose that 
the probability $p$ of transmission 
is $1$. 
To calculate the mean first-encounter time (MFET) between the two walkers, we map their movement
onto that of a single random walker in 2D,
whose allowed positions are represented by the vector $\vect{w}= (w_1, w_2)$; see Fig.~\ref{fig:model}(b). Here, $w_i := W_i - (i-1) \Delta$ is the displacement of walker $i$ from the leftmost site of its habitat $H_{i}$, with $0 \le w_i \le \ell-1$, and 
$\Delta: = \ell-m$ is the distance between the centers of $H_{i}$ and $H_{i+1}$, i.e., the mean spacing between walkers. 

The possible locations where the two walkers may encounter one other then become a sequence of static target positions on the 2D lattice.
When $m=1$,
the overlap region is a single site, 
giving a single target in 2D at 
the bottom right corner, $
(\ell -1,0)$. As $m$ increases, the target set grows, until the limiting case $m = \ell$, when the two walkers occupy the same space and the target set is the main diagonal; see Fig.~\ref{fig:model}(b).

\begin{figure}[tb]
\noindent\includegraphics[scale=0.25]{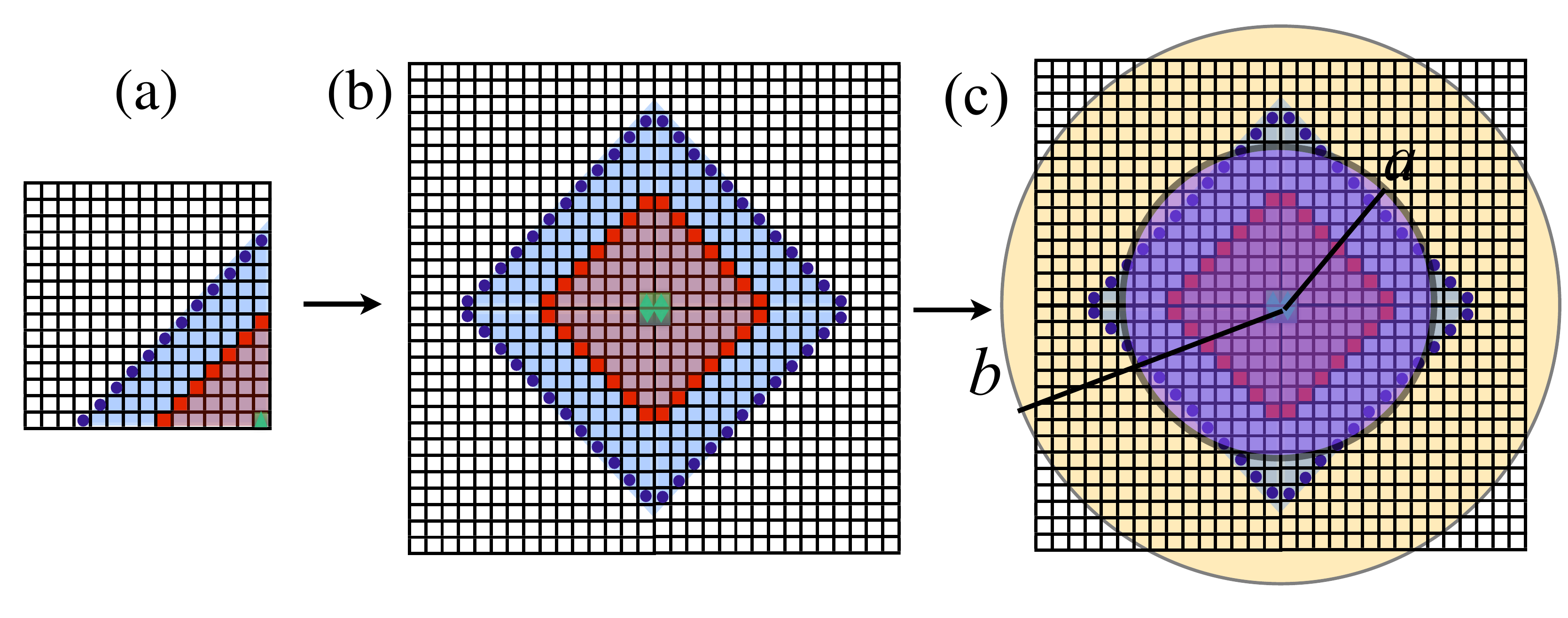}
\caption{(Color online) Unfolding 
for $\ell=15$: (a) original system with reflecting boundaries; (b) unfolded system. Filled, colored sites show targets for the 2D walker for
$m=1$, $m=7$ and $m=12$, shown by triangles, squares, and circles, respectively. 
(c) Approximation for  $m=12$, replacing the region  between the inner rhombus and outer square by an annulus with inner radius $a$ and outer radius $b$.}
\label{fig:twozones}
\end{figure}

The  first-encounter time is now equivalent to the first-passage time to the 2D target set, 
with reflecting boundary conditions at the outer edge of Fig. \ref{fig:model}(b). To compute this, it is 
convenient to consider a symmetrized version of the problem:
instead of reflecting 
a walker which attempts to leave the domain, it is allowed to move unimpeded into a reflected copy of the domain. 
 Doing this for all borders gives rise to an \emph{unfolded} version of the system. This unfolding, a well-known technique in other fields \cite{CM06}, creates an inner absorbing rhombus inside an outer reflecting square as shown in Fig.~\ref{fig:twozones}(a--b).

When $m=1$, 
the extended system  has a ``thick'' target, i.e., four neighboring sites at the center of the lattice, while for other values of $m$ a larger rhombus of target sites is obtained. 
In the continuum limit, we may 
approximate 
the first-encounter distribution of the walkers by computing the first-passage  distribution to the target rhombus.  

When the 2D walker starts inside the rhombus, corresponding to $W_1 > W_2$,
i.e., walker $W_1$ to the right of $W_2$ inside their mutual overlap region,  
one writes the solution of the 2D diffusion equation for the probability distribution $P_{\mathrm{in}}(x,y,t)$ in Cartesian coordinates {with $x$ and $y$ parallel to the edges of the rhombus and} with Dirichlet {(absorbing)} boundary conditions, {since any encounter event corresponds to the arrival at a target, when the movement process terminates}.

The exact solution of this problem with a delta initial condition was obtained in ref.~\cite{tejedoretal2011}, 
in terms of a double infinite series, with each series associated to one of the two orthogonal directions. Uniform initial conditions $P_{\mathrm{in}}(x,y,0)=1/L^{2}$  in our problem, where $L=m \sqrt{2}$,   correspond to integrating Eq.~(12) of ref.~\cite{tejedoretal2011} over the domain. From that expression it is straightforward to calculate the survival probability $S_{\mathrm{in}}(t)= \int _0 ^Ldx \int_0 ^L dy\,P_{\mathrm{in}}(x,y,t)$, and finally the first-passage probability density $F_{\mathrm{in}}(t)=-dS_{\mathrm{in}}(t)/dt$, as
\begin{equation}
F_{\mathrm{in}}(t) = \frac{16}{\pi ^2}\frac{D}{L^2} \sum_{\substack{j \textrm{ odd},\, k \textrm{ odd}}} ^\infty 
\left(\textstyle \frac{1}{j^{2}}+\frac{1}{k^{2}}\right)
e^{-\left(j^{2}+k^{2}\right)\pi^{2}Dt / L^2 }.
\label{fint}
\end{equation}
%
The mean first-encounter time is then $\int_{0}^{\infty} t \, F_{\mathrm{in}}(t) \, dt$, giving
\begin{equation}
\tauin =
m^{2}\left\{\frac{2}{3}-\frac{128}{\pi^{5}}\sum_{k \textrm{ odd }}^{\infty}\frac{\tanh\left[\frac{k\pi}{2}\right]}{k^{5}}\right\} =: C m^{2},
\label{eq:MFETin_numer}
\end{equation}
valid for uniform initial conditions inside the  rhombus, where the diffusion coefficient $D=\frac{1}{4}$, and
so $C \simeq 0.2116$.

When the 2D walker instead starts outside the rhombus, it moves in a much more complicated geometry, and in addition the problem now has mixed boundary conditions. Since this intrinsically two-dimensional eigenvalue problem seems impossible to solve analytically, 
we replace the true geometry by an annulus bounded by two concentric circles, as shown in Fig.~\ref{fig:twozones}(c): one absorbing, of area equal to that of the inner rhombus, and hence radius $a =m \sqrt{2/\pi}$, with  Dirichlet boundary condition {$P_{\mathrm{out}}|_{r=a} =  0$}, {where $r$ is the radial coordinate},  and one reflecting, with area equal to that of the outer square, and so radius $b=2 \ell / \sqrt{\pi}$, with no-flux (Neumann) boundary condition {$\partial P_{\mathrm{out}} / \partial r |_{r=b} = 0$}. 
%

We now seek the radially-symmetric solution  $P_{\mathrm{out}}(r,t)$ of the diffusion equation in this annulus.
{For reasons arising from the epidemic spread problem  below}, we 
take uniform initial conditions inside an annulus $c \leq r \leq b$, for some $c \geq a$.
 Using separation of variables in the radially symmetric diffusion equation, one can determine the probability of occupation $P_{\mathrm{out}}(r,t)$ \cite{crankbook}, the survival probability $S_{\mathrm{out}}(t)=\int_{a}^{b} r \,P_{\mathrm{out}}(r,t) \, dr$, and finally, by differentiation, the 
first-passage probability $F_{\mathrm{out}}(t)=-dS_{\mathrm{out}}(t)/dt$: 
\begin{equation}
F_{\mathrm{out}}(t)=\frac{4D}{b^{2}}\sum_{n=1}^{\infty}G_{n}\left(
z, v
\right)e^{-\alpha_{n}^{2}Dt/b^{2}};
\label{eq:FPTannul}
\end{equation}
with $z:=\frac{a}{b}$, $v:=\frac{a}{c}$ and $z \leq v \leq 1$, and where
\begin{equation}
  G_{n}(z,v) =
\frac
{\rho_{n} \left(\alpha_{n}\frac{z}{v}\right)}
{v\,\,\rho_{n} \left(\alpha_{n}z\right)}
\left \{ \left[ \frac{Y_{0}^{2}\left(\alpha_{n}z\right)}{Y_{1}^{2}\left(\alpha_{n}\right)}-1\right]\left(1-\frac{z^{2}}{v^{2}}\right)
\right \} ^{-1}.
\label{eq:FPTannul2}
\end{equation}
Here, $J_{i}(s)$ and $Y_{i}(s)$ are Bessel functions of order $i$ of the first and second kind, respectively; $\alpha_{n} = \alpha_n (z)$, which depends explicitly on $z$, is the $n$th positive root of the function $\phi_{z}(x) := J_{0}(x\,z)Y_{1}(x) - Y_{0}(x \,z)J_{1}(x)$; and
$\rho_{n}(s):=Y_{1}(s)J_{1}(\alpha_n)-J_{1}(s)Y_{1}(\alpha_n)$.
Note that the roots $\alpha_{n}$ and the functions $\phi_{z}$ and $\rho_{n}$ all depend on the geometry of the annular region, via the parameter $z$.

For the particular case of uniform initial conditions in the whole {annulus, $c=a$, we obtain} 
the 
first-passage time
\begin{equation}
\tauout=\ell^{2}\frac{64}{\pi}\sum_{n=1}^{\infty}\frac{G_{n}\left(
z, 1\right)}{\alpha_{n}^{4}(z)}.
\label{eq:MFETout}
\end{equation}
{Combining 
\eqref{eq:MFETin_numer} and \eqref{eq:MFETout}, we obtain 
the overall MFET: 
}
\begin{equation}
\tauenc = \Pin \tauin +  \Pout \tauout,
\label{eq:generalGFET}
\end{equation}
where $\Pin$ (resp.~$\Pout$) is the probability that the walker starts inside (resp.~outside) the target square.
For {uniform initial conditions 
inside the  whole domain}, 
$\Pin$ is the proportion of area inside the rhombus, $ \frac{m^2}{2\ell^2}$ in the continuum limit, and $\Pout = 1-\Pin$.
We  obtain 
a general expression for $\tauenc / \ell^{2}$, depending
 only on the single dimensionless parameter $\frac{m}{\ell}$.

{When $z=0$, 
the 
walker is in a circular domain, subject only to a reflecting boundary at $r=b$. The  
coefficients $\alpha_{n}$ then reduce to the roots $\beta_{n}$ of $J_{1}$, and it can  be shown that}
for $z \ll 1$ they have the perturbative form $\alpha_{n}(z) \simeq \beta_{n}+\gamma_{n}(z),$ where $\gamma_{n}(z)$ decays to zero slower than any power of $z$. For 
$n=1$, we have
 $\beta_{1}=0$, giving $\alpha_{1}(z)\simeq\sqrt{2/\ln(e^{-3/4}/z)}$ and $G_{1}(z)\simeq [2\ln(e^{-3/4}/z)]^{-1}$, whereas for $n>1$, we 
 have $\beta_{n}>0$
 and $G_{n}(z)\sim \ln^{-2}(z)$, 
  {which comes from} the 
 small-argument expansion of the Bessel function of the second kind of order zero.
For small $z$,  the $n=1$ term
thus increases logarithmically, while the $n>1$ terms decay to zero. 

We thus obtain the following  approximation:
\begin{equation}
\frac{\tauenc}{\ell^2} \simeq \frac{C}{2} \frac{m^4}{\ell^4} + \frac{8}{\pi}\ln\left(\frac{e^{-3/4}\sqrt{2}\ell}{\,m}\right),
\label{eq:approxGFET}
\end{equation}
which exhibits the well-known logarithmic dependence of mean first-passage times for small target size \cite{rednerbook2001,condaminetal2007}. 
Note that from the exact first-passage distribution \eqref{eq:FPTannul} to a circular target, we can determine all higher-order moments of the first-encounter time,
{as will be reported in detail elsewhere}.

Numerically-exact values were  obtained by solving  recurrence relations for the mean first-passage time $\bar{\tau}(x,y)$ to  any target site starting from $(x,y)$: 
$\bar{\tau}(x,y) = 1 + \frac{1}{4} [ \bar{\tau}(x-1, y) + \bar{\tau}(x+1, y) + \bar{\tau}(x, y-1) + \bar{\tau}(x, y+1) ] $.
Target sites have $\bar{\tau}(x,y) = 0$, and the reflecting boundary conditions must be taken into account.
This gives a sparse system of linear equations, which may be solved by suitable numerical methods. The results are then averaged over all sites $(x,y)$ to give $\tauenc$.

\begin{figure}[t]
\includegraphics[scale=0.325]{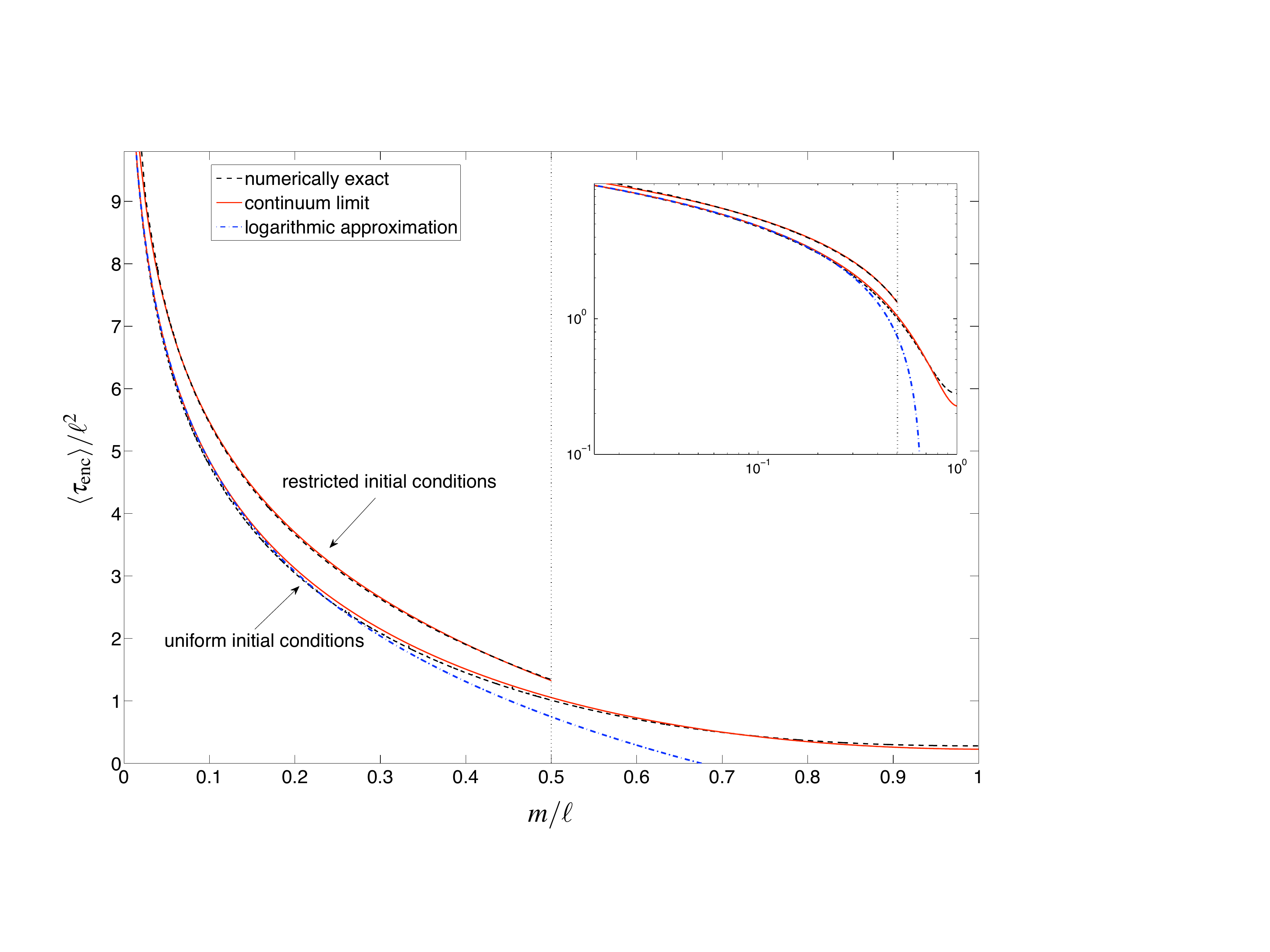}
	\caption{(Color online) Dependence of $\tauenc / \ell^2$ as a function of $m/\ell$. Numerical results are shown
with (black) dashed lines, the analytical continuum limits (\ref{eq:generalGFET}) with a (red) solid line, and the approximate expression (\ref{eq:approxGFET}) with  a (blue) dash-dotted line. The inset shows the corresponding curves on a log--log plot. The mean first-encounter time  relevant to epidemic propagation (walker $W_{1}$  uniform in its left overlap region, i.e., restricted initial conditions) is shown only for $m / \ell < \frac{1}{2}$, delimited by a vertical dotted line, since we  consider only cases for which individuals always maintain an exclusive area. 
	}
	\label{fig:collapse}
\end{figure}

Figure~\ref{fig:collapse} shows such numerically-exact results for several combinations of $m$ and $\ell$, which all collapse onto a single curve;
 this, in turn, is very well approximated by the analytic expression
\eqref{eq:generalGFET} for the continuum limit. We also compare  the first-order approximation  \eqref{eq:approxGFET}, 
which mirrors the analytical results for small $m /\ell$, but is not accurate for $m / \ell \gtrsim 0.3$. 

\paragraph{Epidemic 
spread:-}
To pass from microscopic interactions between  walkers to  macroscopic  epidemic propagation, we consider a  chain   
of habitats $H_i$,  with $i=0$, $1$, \ldots, each {containing} one confined walker, as {sketched}
in Fig.~\ref{fig:model}.
Starting with the left walker $0$ infected, the time to infect walker $n$ 
is 
$\tau^{(n)}$ =
 $\sum_{i=1}^n \tau_i$, where the $\tau_i$ are independent random variables giving the times for walker $i-1$ 
to infect walker $i$.

Each pair of walkers $i-1$ and $i$ is an identical copy of the same system (except the leftmost pair).
However,  the initial conditions are no longer uniform over the entire square, since
 when $W_{i-1}$ infects its neighbor $W_{i}$, the initial position of the latter must lie  inside the corresponding overlap region $R_{i-1}$.
%

For the 2D walker, this  corresponds to taking a distribution that is uniform only  within two vertical strips, of width $m$ and height $2 \ell$, at the left and right edges of the unfolded square domain shown in Fig.~\ref{fig:twozones}(b). 
By symmetry, the same result is  obtained starting from two horizontal strips, so it can be  approximated by a  square strip of width $m$ around all edges of the domain, neglecting double counting in the corners where the horizontal and vertical strips overlap. In turn, we approximate this   by the annulus $c \le r \le b$, where $c = 2\ell (1 - m / \ell ) / \sqrt{\pi}$, with area equal to that of the square strip.

Taking uniform initial conditions in this annulus,with
$\Pin=0$ in Eq.~\eqref{eq:generalGFET},  gives the MFET $\tauenc = \tauenc_{\mathrm{out}}$, whose expression is as in  Eq.~\eqref{eq:MFETout}, but with $G_{n}(z,1)$ replaced by $G_{n}(z,v)$, with $v = m / [\sqrt{2} \ell (1 - m/\ell)]$.
{This  is  compared to numerically-exact values in Fig.~\ref{fig:collapse}, with excellent agreement.} We remark that the relevant regime for territorial animals is $m/\ell<1/2$, which corresponds to the presence of an exclusive core area.

{If the infection is  transmitted 
with probability $p<1$ at an encounter,}
then the walkers may meet repeatedly before  the infection is transmitted, at any position in the overlap region.
Returns to the target set, in which the 2D walker starts from one of the possible targets and comes back to hit any of them, must then be accounted for.  
The infection is transmitted at the $k$th encounter with probability $p_{k} := p(1-p)^{k-1}$, {giving the mean infection time 
$\mean{\tau} = \sum_{k=1}^{\infty} p_{k} \mean{\tau | k},$ where $\mean{\tau | k}$ is the 
mean infection time \emph{conditional}} on  infection occurring at the $k$th encounter. We have $\mean{\tau | k} = \tauenc + (k-1) \tauret$,
where $\tauret$ is the mean return time,
giving
\begin{equation}
\meantau = \tauenc + (\textstyle \frac{1}{p} - 1) \tauret.
\label{eq:effect-of-p}
\end{equation}

The mean return time $\tauret$ to a set $U$ can  be calculated for discrete stochastic models, using the Kac recurrence lemma \cite{KacRecurrenceThmBullAMS1947, sanders2009}, as
$\tauret = |\Omega| / |U|$, where $\Omega$ is the set of all possible configurations and $|\cdot|$ denotes the 
 number of configurations in a set. 
Taking $U$ as the target set, 
we have
 $|U| = m$, giving $\tauret / \ell^{2} = 1 / m$,  {no longer} a function of the dimensionless parameter $m/\ell$. 
{Eq.~\eqref{eq:effect-of-p} then gives the mean infection time $\meantau$ for any $p \leq 1$.}

{The previous results may now be combined to calculate the speed of } propagation of an epidemic  in the chain of habitats.
We 
{call $\omega(t)$ the position of the epidemic front, i.e., of the right-most infected walker }
\cite{sokolov_epjb}.
The $n$th walker is infected at time $\tau^{(n)}$, at which time the front is at $\omega(\tau^{(n)}) = n (\ell-m)$.
Averaging over realizations,  the front propagation is  linear in $n$, with mean speed
$ (\ell - m) / \mean{\tau}$. 

\begin{figure}[t]
\noindent\includegraphics[scale=0.46]{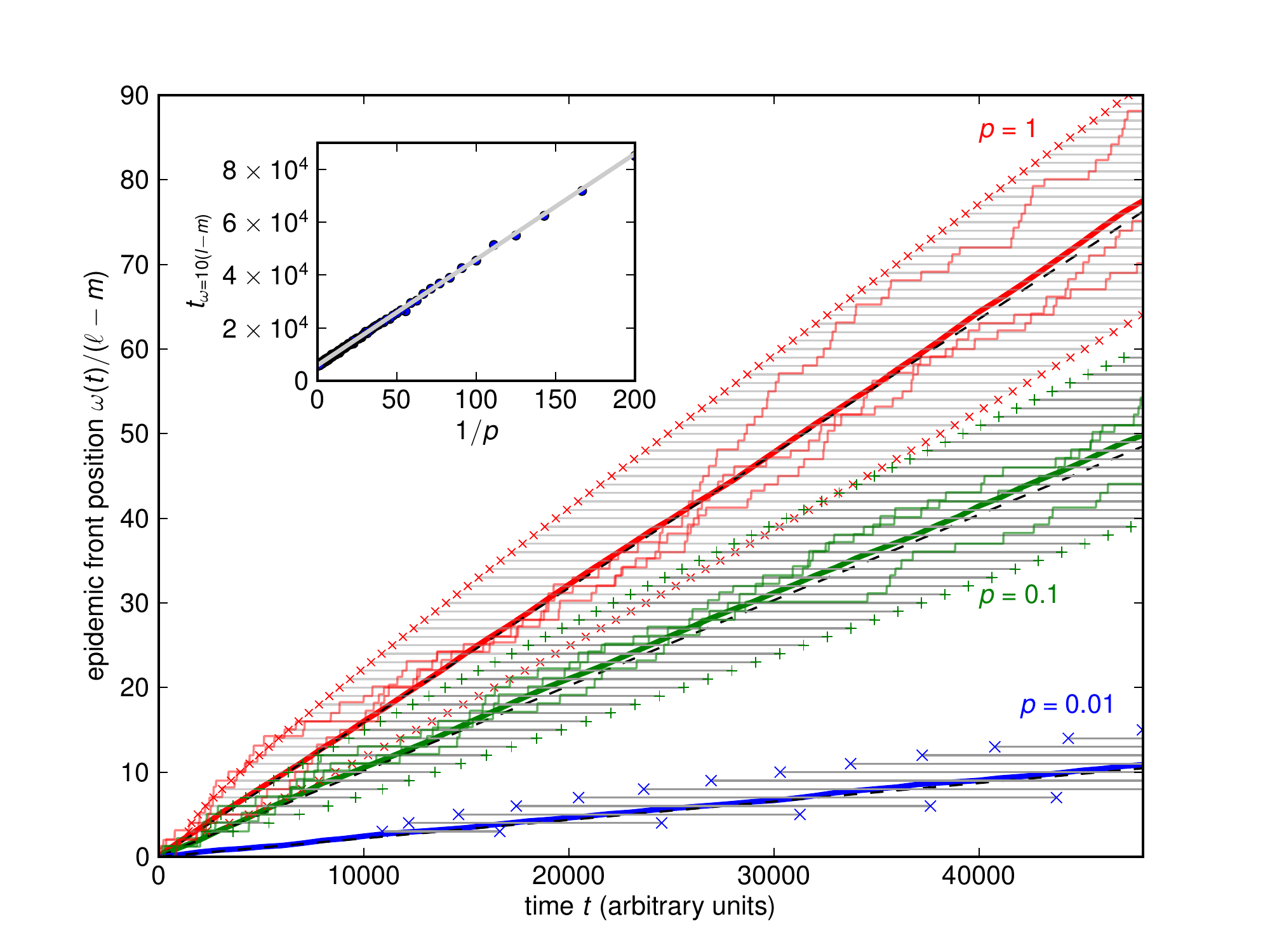}
\caption{(Color online) Analytical (dashed lines) and numerical (thick lines, average over $100$ realizations) position 
$\omega (t)$ of the epidemic front for $\ell=20$ and $m=5$ and
  different transmission probabilities $p$, as a function of time $t$ (number of simulation steps);
 $\tauenc \simeq 629$ for these values of $\ell$ and $m$.  
 For illustrative purposes three realizations are shown for $p=1$ (red online) and $p=0.1$ (green). 
 Crosses and  gray horizontal bars show the range of possible times to arrive at a given front position, predicted by the law of the iterated logarithm. Inset:  mean time to arrive at the front position $10(\ell - m)$ as a function of $1/p$, comparing analytical (solid line) and numerical results (dots; averaged over $200$ realizations).
\label{fig:propagation}
}
\end{figure}

  Figure~\ref{fig:propagation} compares the analytical result for the mean front position 
  with numerical simulations, showing excellent agreement for different transmission probabilities $p$. Trajectories of individual realizations are also shown, as well as  
    ranges of the possible times to arrive at a given  front position, given by the 
law of the iterated logarithm \cite{grimmettstirzaker}, which states that the infection time of walker $n$ {lies} within the interval $n \mean{\tau} \pm \sqrt{2 n \ln (\ln n)} \, \sigma(\tau)$ for large enough $n$. Here, $\sigma(\tau)$ denotes the standard deviation of the first-encounter time, 
{calculated} analytically from the exact expression \eqref{eq:FPTannul} for $F_{\mathrm{out}}$ for $p=1$, and numerically for the other values of $p$. This agrees with the several trajectories shown for  $p=1$ and $p=0.1$, which are contained inside these bounds {for large times}.

For two-dimensional domains, this approach  can be extended by
looking for encounters along one axis, and conditioning on the walkers' positions being identical also along the other axis. Preliminary results suggest different scalings as a function of domain size, depending on the form of the overlap region; this extension will form part of a future publication.

Our framework can also be extended to %
include 
spatial dependence of the transmission probability, due to heterogeneity in neighboring pairs, e.g., young versus adult individuals.
 The infection front then moves in a medium with infection rates which are not all equal (disordered); 
 effective-medium theory \cite{kenkreetal2009}
  provides the tools 
  to 
  calculate the propagation speed 
in terms of the
 {transmission   distribution}.

 The authors thank D.~Boyer, N.~Kenkre and H.~Larralde for helpful discussions.
 Financial support from EPSRC grant EP/I013717/1 and  DGAPA-UNAM PAPIIT grant 
IN116212 is acknowledged, and SP acknowledges a studentship from CONACYT, Mexico.

\bibliographystyle{apsrev}
\bibliography{Biblio_caminantes2}

\end{document}